\documentclass{article}
\usepackage[left=1.125in,right=1.125in,
            top=1.125in,bottom=1.125in,footskip=.35in]{geometry}

\usepackage{setspace}
\usepackage{graphicx}
\usepackage{float}
\usepackage{bm}
\usepackage{physics}
\usepackage{braket}
\usepackage{xcolor}

\title{Quantum simulations of materials on near-term quantum computers} 

\author
{He Ma,$^{1, 3}$ Marco Govoni$^{2, 3}$, Giulia Galli$^{1, 2, 3\ast}$
\\
\normalsize{$^{1}$Department of Chemistry, University of Chicago, Chicago, IL 60637, USA}\\
\normalsize{$^{2}$Pritzker School of Molecular Engineering, University of Chicago, Chicago, IL 60637, USA}\\
\normalsize{$^{3}$Materials Science Division and Center for Molecular Engineering,}\\
\normalsize{Argonne National Laboratory, Lemont, IL 60439, USA}\\
\normalsize{$^\ast$Corresponding author. Email: 
gagalli@uchicago.edu}
}

\date{}

\begin{document}

\maketitle

\begin{abstract}
Quantum computers hold promise to enable efficient simulations of the properties of molecules and materials; however, at present they only permit \textit{ab initio} calculations of a few atoms, due to a limited number of qubits. In order to harness the power of near-term quantum computers for simulations of larger systems, it is desirable to develop hybrid quantum-classical methods where the quantum computation is restricted to a small portion of the system.  This is of particular relevance for molecules and solids  where an active region requires a higher level
of theoretical accuracy than its environment. Here we present a quantum embedding theory for the calculation of strongly-correlated electronic states of active regions, with the rest of the system described within density functional theory. We demonstrate the accuracy and effectiveness of the approach by investigating several defect quantum bits in semiconductors that are of great interest for quantum information technologies. We perform calculations on quantum computers and show that they yield results in agreement with those obtained with exact diagonalization on classical architectures, paving the way to simulations of realistic materials on near-term quantum computers.

\end{abstract}
\pagebreak

\section*{Introduction}

In the last three decades, atomistic simulations based on the solution of the basic equation of quantum mechanics have played an increasingly important role in predicting the properties of functional materials, encompassing catalysts and energy storage systems for energy applications, and materials for quantum information science. Especially in the case of complex, heterogeneous materials, the great majority of first-principles simulations are conducted using density functional theory (DFT), which is in principle exact but in practice requires approximations to enable calculations. Within its various approximations, DFT has been extremely successful in predicting numerous properties of solids, liquids and molecules, and in providing key interpretations to a variety of experimental results; however it is often inadequate to describe so-called strongly-correlated electronic states \cite{Cohen2008,Su2018}. We will use here the intuitive notion of strong correlation as pertaining to electronic states that cannot be described by mean-field theories. Several theoretical and computational methods have been developed over the years to treat systems exhibiting strongly-correlated electronic states, including dynamical mean-field theory \cite{Georges1996,Kotliar2006} and quantum Monte-Carlo \cite{Ceperley1986,Wagner2016}; in addition, \textit{ab initio} quantum chemistry methods, traditionally developed for molecules, have been recently applied to solid state problems as well \cite{Sun2017}. Unfortunately, these approaches are computationally demanding and it is still challenging to apply them to complex materials containing defects and interfaces, even using high-performance computing architectures.

Quantum computers hold promise to enable efficient quantum mechanical simulations of weakly and strongly-correlated molecules and materials alike \cite{Aspuru-Guzik2005,Bravyi2017,Babbush2018,Kivlichan2018,Motta2019,Ollitrault2019,Cao2019,Bauer2020}; in particular when using quantum computers, one is able to efficiently simulate systems of interacting electrons, with an exponential speedup over exact diagonalization on classical computers (full configuration interaction, FCI, calculations). Thanks to decades of successful experimental efforts, we are now entering the noisy intermediate-scale quantum (NISQ) era \cite{Preskill2018}, with quantum computers expected to have on the order of 100 quantum bits (qubits); unfortunately this limited number of qubits still prevents straightforward quantum simulations of realistic molecules and materials, whose description requires hundreds of atoms and thousands to millions of degrees of freedom to represent the electronic wavefunctions. An important requirement to tackle complex chemistry and material science problems using NISQ computers is the reduction of the number of electrons treated explicitly at the highest level of accuracy \cite{Rubin2016,Yamazaki2018}. For instance, building on the idea underpinning dynamical mean field theory \cite{Georges1996,Kotliar2006}, one may simplify complex molecular and material science problems by defining active regions (or building blocks) with strongly-correlated electronic states, embedded in an environment that may be described within mean-field theory \cite{Bauer2016,Kreula2016,Rungger2019}.

In this work, we present a quantum embedding theory built on DFT, which is scalable to large systems and which includes the effect of exchange-correlation interactions of the environment on active regions, thus going beyond commonly adopted approximations. In order to demonstrate the effectiveness and accuracy of the theory, we compute ground and excited state properties of several spin-defects in solids including the negatively charged nitrogen-vacancy (NV) center \cite{Davies1976,Rogers2008,Doherty2011,Maze2011,Choi2012,Doherty2013,Goldman2015}, the neutral silicon-vacancy (SiV) center \cite{Haenens2011,Gali2013,Green2017,Rose2018,Green2019,Thiering2019} in diamond, and the Cr impurity (4+) in 4H-SiC \cite{Son1999,Koehl2017,Diler2019}. These spin-defects are promising platforms for solid-state quantum information technologies, and they exhibit strongly-correlated electronic states that are critical for the initialization and read-out of their spin states \cite{Weber2010,Seo2016,Seo2017,Ivady2018,Dreyer2018,Anderson2019}. Our quantum embedding theory yields results in good agreement with existing measurements. In addition, we present the first theoretical predictions for the position and ordering of the singlet states of SiV and of Cr, and we provide an interpretation of experiments which have so far remained unexplained.

Importantly, we report calculations of spin-defects using a quantum computer \cite{Qiskit,Yorktown}. Based on the effective Hamiltonian derived from the quantum embedding theory, we investigated the strongly-correlated electronic states of the NV center in diamond using quantum phase estimation algorithm (PEA) \cite{Abrams1997,Aspuru-Guzik2005} and variational quantum eigensolvers (VQE) \cite{Peruzzo2014,McClean2016,Kandala2017}, and we show that quantum simulations yield results in agreement with those obtained with classical FCI calculations. Our findings pave the way to the use of near term quantum computers to investigate the properties of realistic heterogeneous materials with first-principles theories.

\hspace{1cm}

\section*{Results}

\noindent\textbf{General strategy}

\begin{figure}[H]
  \includegraphics[width=6in]{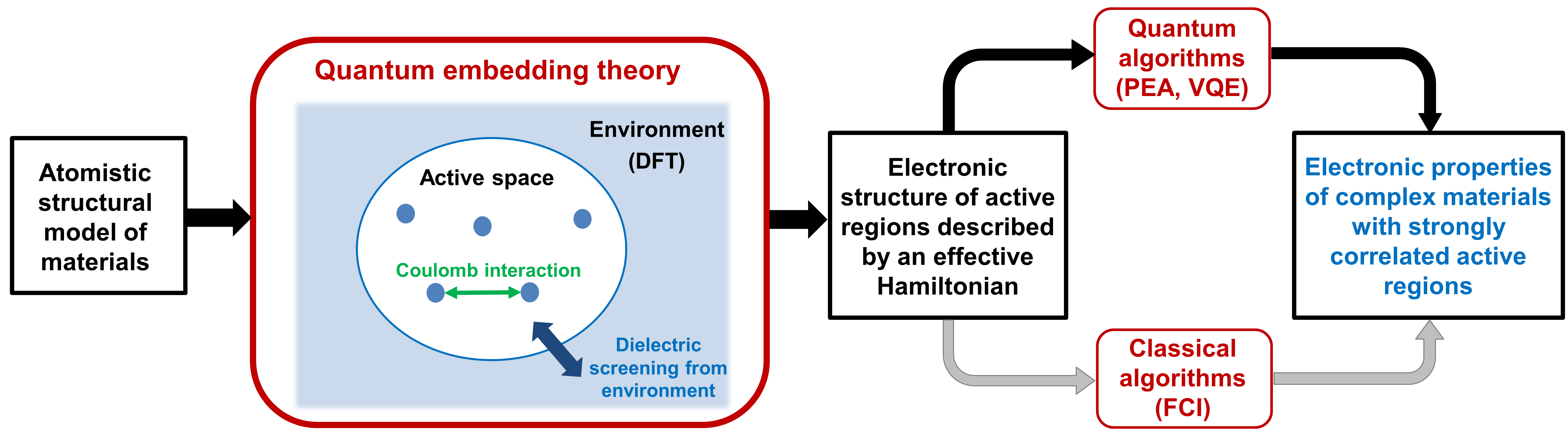}
  \caption{General strategy for quantum simulations of materials using quantum embedding. The full system is separated into an active space and its environment, with the electronic states in the active space described by an effective Hamiltonian solved with either classical (e.g. full configuration interaction, FCI) or quantum algorithms (e.g. phase estimation algorithm (PEA), variational quantum eigensolver (VQE)). The effective interaction between electrons in the active space includes the bare Coulomb interaction and a polarization term arising from the dielectric screening of the environment (see text), which is evaluated including exchange-correlation interactions.}
  \label{strategy}
\end{figure}

\noindent We summarize our strategy in Fig. \ref{strategy}. Starting from an atomistic structural model of materials (e.g. obtained from DFT calculations or molecular dynamics simulations), we identify active regions with strongly-correlated electrons, which we describe with an effective Hamiltonian that includes the effect of the environment on the active region. This effective Hamiltonian is constructed using the quantum embedding theory described below, and its eigenvalues can be obtained by either classical algorithms such as exact diagonalization (FCI) or quantum algorithms.

\hspace{1cm}

\noindent\textbf{Embedding Theory}

\noindent A number of interesting quantum embedding theories have been proposed over the past decades \cite{Sun2016}. For instance, density functional embedding theory has been developed to improve the accuracy and scalability of  DFT calculations \cite{Huang2011,Goodpaster2014,Jacob2014,Genova2014,Wen2019}.  Density matrix embedding theory (DMET) \cite{Knizia2012,Wouters2016,Pham2019} and various Green's function based approaches \cite{Nguyen2016,Dvorak2019}, e.g. dynamical mean field theory (DMFT), have been developed to describe systems with strongly-correlated electronic states. At present, \textit{ab initio} calculations of materials using DMET and DMFT have been limited to relatively small unit cells (a few tens of atoms) of {\it pristine} crystals, due to their high computational cost \cite{Zhu2019,Cui2019}. In this work, we present a quantum embedding theory that is applicable to strongly-correlated electronic states in realistic {\it heterogeneous} materials and we apply it to systems with hundreds of atoms. The theory, inspired by the constrained random phase approximation (cRPA) approach \cite{Aryasetiawan2004,Miyake2009,Hirayama2017}, does not require the explicit evaluation of virtual electronic states \cite{Wilson2008,Govoni2015}, thus making the method scalable to materials containing thousands of electrons. Furthermore, cRPA approaches contain a specific approximation (RPA) to the screened Coulomb interaction, which neglects exchange-correlation effects and may lead to inaccuracies in the description of dielectric screening. Our embedding theory goes beyond the RPA by explicitly including exchange-correlation effects, which are evaluated with a recently developed finite-field algorithm \cite{Ma2018, Nguyen2019}.

The embedding theory developed here aims at constructing an effective Hamiltonian operating on an active space (A), defined as a subspace of the single-particle Hilbert space:
\begin{equation} \label{Heff}
    H^{\mathrm{eff}} = \sum_{ij}^{\mathrm{A}} t^{\mathrm{eff}}_{ij} a_{i}^{\dagger} a_{j} + \frac{1}{2} \sum_{ijkl}^{\mathrm{A}} V^{\mathrm{eff}}_{ijkl} a_{i}^{\dagger} a_{j}^{\dagger} a_{l} a_{k}.
\end{equation}
Here $t^{\mathrm{eff}}$ and $V^{\mathrm{eff}}$ are one-body and two-body interaction terms that take into account the effect of all the electrons that are part of the environment (E) in a mean-field fashion, at the DFT level. An active space can be defined, for example, by solving the Kohn-Sham equations of the full system and selecting a subset of eigenstates among which electronic excitations of interest take place (e.g. defect states within the gap of a semiconductor or insulator). To derive an expression for $V^{\mathrm{eff}}$ that properly accounts for all effects of the environment including exchange and correlation interactions, we define the environment density response function (reducible polarizability) $\chi^\mathrm{E} = \chi_0^E + \chi_0^\mathrm{E} f \chi^E$, where $\chi_0^\mathrm{E} = \chi_0 - \chi_0^\mathrm{A}$ is the difference between the polarizability of the Kohn-Sham system $\chi_0$ and its projection onto the active space $\chi_0^\mathrm{A}$ (see Supplementary Information (SI)). $\chi^\mathrm{E}$ thus represents the density response outside the active space. The term $f = V + f_{\mathrm{xc}}$ is often called the Hartree-exchange-correlation kernel, where $V$ is the Coulomb interaction and the exchange-correlation kernel $f_{\mathrm{xc}}$ is defined as the derivative of the exchange-correlation potential with respect to the electron density. We define the effective interactions between electrons in A as
\begin{equation} \label{Veff}
    V^{\mathrm{eff}} = V + f \chi^\mathrm{E} f,
\end{equation}
given by the sum of the bare Coulomb potential and a polarization term arising from the density response in the environment E. When the RPA is adopted, the exchange-correlation kernel $f_{\mathrm{xc}}$ is neglected in Eq. \ref{Veff} and the expression derived here reduces to that used within cRPA. We represent $\chi^\mathrm{E}$ and $f$ on a compact basis obtained from a low-rank decomposition of the dielectric matrix \cite{Wilson2008,Govoni2015} that allows us to avoid the evaluation and summation over virtual electronic states. Once $V^{\mathrm{eff}}$ is defined, the one-body term $t^{\mathrm{eff}}$ can be computed by subtracting from the Kohn-Sham Hamiltonian a term that accounts for Hartree and exchange-correlation effects in the active space (see SI). 

\hspace{1cm}

\noindent\textbf{Embedding theory applied to spin-defects}

\noindent The embedding theory presented above is general and can be applied to a variety of systems for which active regions, or building blocks, with strongly-correlated electronic states may be identified: for example active sites in inorganic catalysts or organic molecules or defects in solids and liquids (e.g. solvated ions in water). Here we apply the theory to spin-defects including NV and SiV in diamond and Cr in 4H-SiC. Most of these defects' excited states are strongly-correlated (they cannot be represented by a single Slater determinant of single-particle orbitals), as shown e.g. for the NV center in diamond by Bockstedte et al.\cite{Bockstedte2018}. We demonstrate that our embedding theory can successfully describe the many-body electronic structure of different types of defects including transition metal atoms; our results not only confirm existing experimental observations but also provide a detailed description of the electronic structure of defects not presented before, which sheds light into their optical cycles.

\begin{figure}[H]
  \centering
  \includegraphics[width=5.5in]{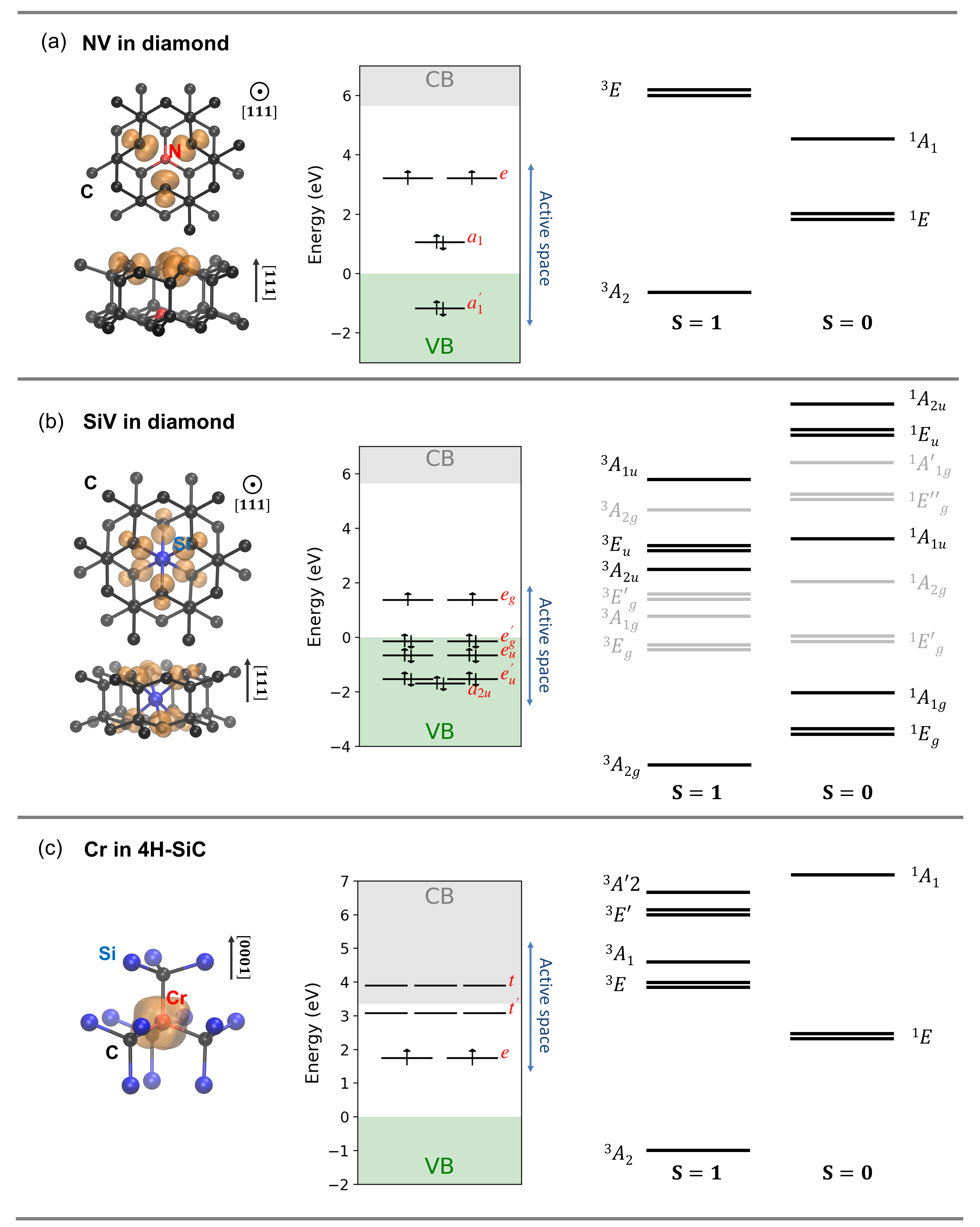}
  \caption{Electronic structure of the negatively-charged nitrogen vacancy (NV) in diamond (a), the neutral silicon vacancy (SiV) in diamond (b) and the Cr impurity (4+) in 4H-SiC (c). Left panels show spin densities obtained from spin unrestricted DFT calculations. Middle panels show the position of single-particle defect levels computed by spin restricted DFT calculations. States included in active spaces (see text) are indicated by blue vertical lines. Right panels show the symmetry and ordering of the low-lying many-body electronic states obtained by exact diagonalization (FCI calculations) of effective Hamiltonians constructed with exchange-correlation interactions included.}
  \label{defects}
\end{figure}

We first performed spin-restricted DFT calculations using hybrid functionals \cite{Skone2014} to obtain a mean-field description of the defects and of the whole host solid. The spin restriction ensures that both spin channels are treated on an equal footing and that there is no spin-contamination when building effective Hamiltonians. Based on our DFT results, we then selected active spaces that include single-particle defect wavefunctions and relevant resonant and band edge states. We verified that the size of the chosen active spaces yields converged excitation energies (see SI). We then constructed effective Hamiltonians (Eq. \ref{Heff}-\ref{Veff}) by taking into account exchange-correlation effects, and we obtained many-body ground and excited states using classical (FCI) and, for selected cases, quantum algorithms (PEA, VQE). All calculations were performed at the spin triplet ground state geometries obtained by spin-unrestricted DFT calculations, thus obtaining vertical excitation energies (equal to the sums of zero phonon line (ZPL) and Stokes energies). It is straightforward to extend the current approach to compute potential energy surfaces at additional geometries, so as to include relaxations and Jahn-Teller effects \cite{Bockstedte2018,Thiering2019}. In Fig. \ref{defects} we present atomistic structures, single-particle defect levels and the many-body electronic structure of three spin-defects. Several relevant vertical excitation energies are reported in Table \ref{excitation_energies}, and additional ones are given in the SI. In the following discussion, lower-case symbols represent single-particle states obtained from DFT and upper-case symbols represent many-body states.

For the NV in diamond, we constructed effective Hamiltonians (Eq. \ref{Heff}) by using an active space that includes $a_1$ and $e$ single-particle defect levels in the band gap and states near the valence band maximum (VBM). Our FCI calculations correctly yield the symmetry and ordering of the low-lying $^3A_2$, $^3E$, $^1E$ and $^1A_1$ states. The vertical excitation energies reported in Table 1 show that including exchange-correlation effects yields results in better agreement with experiments than those obtained within the RPA. 

In the case of the SiV in diamond, we built effective Hamiltonians using an active space with the $e_u$ and $e_g$ defect levels and states near the VBM, including resonant $e'_u$ and $e'_g$ states. Effective Hamiltonians including or neglecting exchange-correlation effects yield similar results, with the excitation energies obtained beyond RPA being slightly higher. We predicted the first optically-allowed excited state to be a $^3E_{u}$ state with vertical excitation energy of 1.59 eV, in good agreement with the sum of 1.31 eV ZPL measured experimentally \cite{Haenens2011} and 0.258 eV Stokes shift estimated using an electron-phonon model \cite{Thiering2019}. Remarkably, our calculations predicted a $^3A_{2u}$ state 11 meV below the $^3E_{u}$ state, supporting a recent experimental observation by Green et al. \cite{Green2019}, which proposed a $^3A_{2u}$-$^3E_{u}$ manifold with 7 meV separation in energy. The small difference in energy splitting between our results and experiment is likely due to geometry relaxation effects not yet taken into account in our study. In addition to states of $u$ symmetry generated by $e_u \rightarrow e_g$ excitations, we observed a number of optically dark states of $g$ symmetry (grey levels in Fig. \ref{defects}b) originating from the excitation from the $e'_g$ level and the VBM states to the $e_g$ level. 

Despite significant efforts \cite{Green2017,Rose2018,Green2019,Thiering2019}, several important questions on the singlet states of SiV remain open. These states are crucial for a complete understanding of the optical cycle of the SiV center. Our predicted ordering of singlet states of SiV is shown in Fig. \ref{defects}b. We find the vertical excitation energies of the $^1A_{1u}$ state to be slightly higher than that of the $^3A_{2u}$-$^3E_{u}$ triplet manifold, suggesting that the intersystem crossing (ISC) from $^3A_{2u}$ or $^3E_{u}$ to singlet states may be energetically unfavorable (first-order ISC to lower $^1E_{g}$ and $^1A_{1g}$ states are forbidden). We note that the $^1E_{u}$ and $^1A_{2u}$ states are much higher in energy than $^1A_{1u}$ and are not expected to play a significant role in the optical cycle. In addition the first-order ISC process from the lowest energy singlet state $^1E_{g}$ to the $^3A_{2g}$ ground state is forbidden by symmetry. Overall our results indicate that the $^3A_{2g}$ state is populated through higher-order processes and therefore the spin-selectivity of the full optical cycle is expected to be low. While more detailed studies including spin-orbit coupling are required for definitive  conclusions, our predictions shed light on the strongly-correlated singlet states of SiV and provide a possible explanation for the experimental difficulties in measuring optically-detected magnetic resonance (ODMR) of SiV.

We now turn to Cr in 4H-SiC, where we considered the hexagonal configuration. We constructed effective Hamiltonians with the half-filling $e$ level in the band gap and states near the conduction band minimum (CBM) including resonance states. Upon solving the effective Hamiltonian, we predict the lowest excited state to be a $^1E$ state arising from $e \rightarrow e$ spin-flip transition, with excitation energy of 1.09 (0.86) eV based on embedding calculations beyond (within) the RPA. Results including exchange-correlation effects are in better agreement with the measured ZPL of 1.19 eV \cite{Son1999}, where the Stokes energy is expected to be small given the large Debye-Waller factor\cite{Diler2019}. There is currently no experimental report for the triplet excitation energies of Cr in 4H-SiC, but our results are in good agreement with existing experimental measurements for Cr in GaN, a host material with a crystal field strength similar to that of 4H-SiC \cite{Koehl2017}. We predict the existence of a $^3E+{}^3A_1$ manifold at $\simeq$ 1.4 eV and a $^3E'+{}^3A'_2$ manifold at $\simeq$ 1.7 eV above the ground state (Fig. \ref{defects}c), resembling the $^3T_2$ manifold (1.2 eV) and $^3T_1$ manifold (1.6 eV) for Cr in GaN observed experimentally \cite{Heitz1995}. We note that in many cases it is challenging to study materials containing transition metal elements with DFT \cite{Anisimov1997}. The agreement between FCI results and experimental measurements clearly demonstrates that the embedding theory developed here can effectively describe the strongly-correlated part of the system, while yielding at the same time a quantitatively correct description of the environment. 

\hspace{1cm}

\noindent\textbf{Quantum simulations of spin-defects}

\begin{figure}[H]
  \centering
  \includegraphics[width=3.5in]{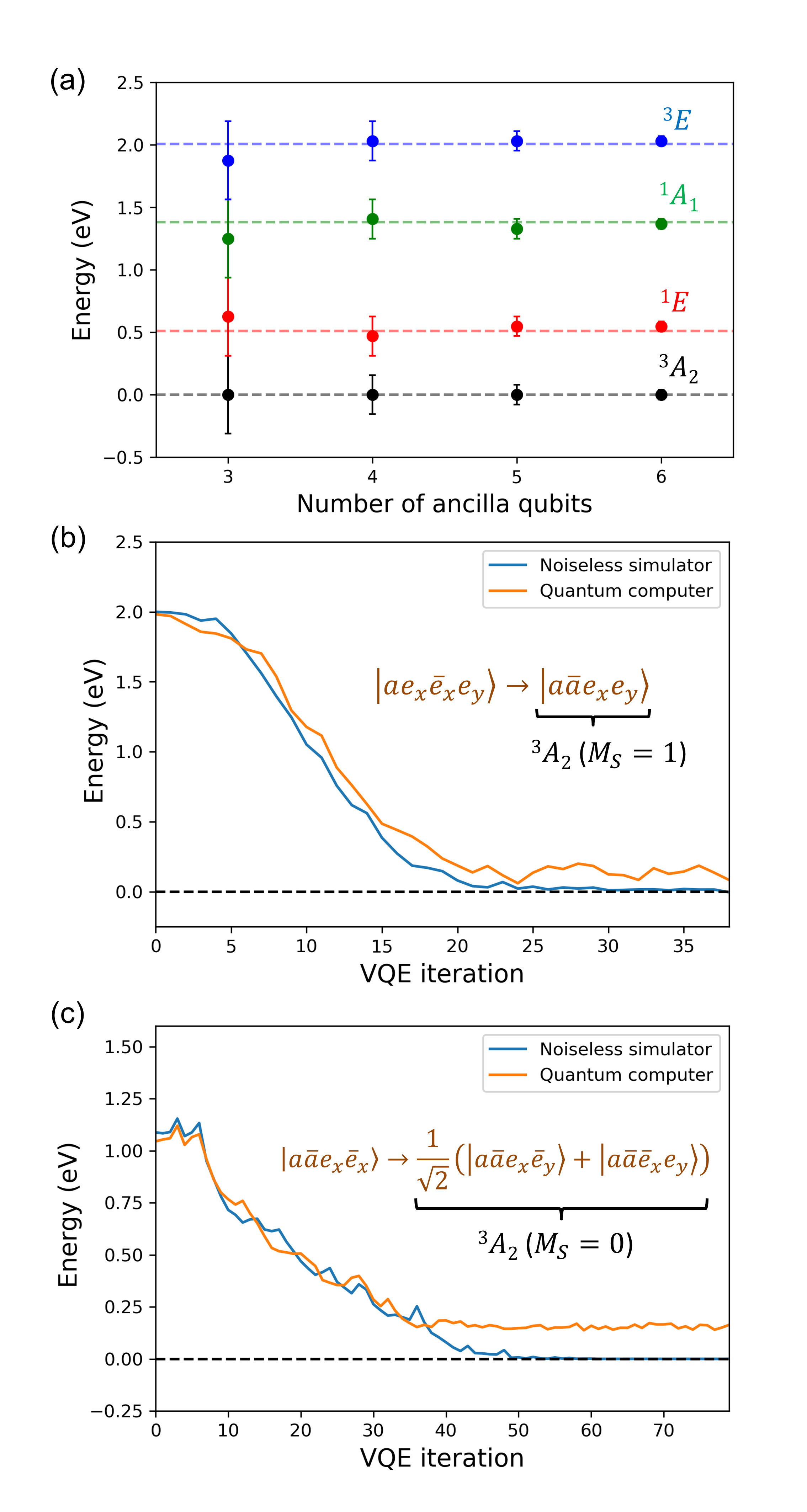}
  \caption{Quantum simulations of a minimum model of the NV center in diamond using the phase estimation algorithm (PEA) and a variational quantum eigensolver (VQE). The energy of the $^3A_2$ ground state manifold is set to zero for convenience. (a) PEA estimation of ground and excited states of the NV center. Error bars represent the uncertainties due to the finite number of ancilla qubits used in the simulations; dashed lines show classical FCI results. (b) VQE estimation of ground state energy, starting from $\ket{a e_x \bar{e}_x e_y}$ state ($M_S = 1$, see text). (c) VQE estimation of ground state energy, starting from $\ket{a \bar{a} e_x \bar{e}_x}$ state ($M_S = 0$); strongly-correlated $\frac{1}{2}\left( \ket{a \bar{a} e_x \bar{e}_y} + \ket{a \bar{a} \bar{e}_x e_y} \right)$ state ($M_S = 0$ state in the $^3A_2$ manifold) is obtained with VQE. }
  \label{quantum}
\end{figure}

\noindent The results presented in the previous section were obtained using classical algorithms. We now turn to the use of quantum algorithms. To perform quantum simulations with PEA and VQE, we constructed a minimum model of an NV center including only $a_1$ and $e$ orbitals in the band gap. This model (4 electrons in 6 spin orbitals) yields excitation energies within 0.2 eV of the converged results using a larger active space. In Fig. \ref{quantum} we show the results of quantum simulations.

We first performed PEA simulations with a quantum simulator (without noise) \cite{Qiskit} to compute the energy of $^3A_2$, $^3E$, $^1E$ and $^1A_1$ states. We used molecular orbital approximations of these states derived from group theory \cite{Doherty2011} as initial states for PEA, which are single Slater determinant for $^3A_2$ ($M_S = 1$) and $^3E$ ($M_S = 1$) states, and superpositions of two Slater determinants for $^1E$ and $^1A_1$ states. As shown in Fig. \ref{quantum}a, PEA results converge to classical FCI results with an increasing number of ancilla qubits.

We then performed VQE simulations with a quantum simulator and with the IBM Q 5 Yorktown quantum computer \cite{Yorktown}. We estimated the energy of the $^3A_2$ ground state manifold by performing VQE calculations for both the single-Slater-determinant $M_{S} = 1$ component and the strongly-correlated $M_{S} = 0$ component. Within a molecular orbital notation, $M_{S} = 1$ and $M_{S} = 0$ ground states can be represented as $\ket{a \bar{a} e_x e_y}$ and $\frac{1}{2}\left( \ket{a \bar{a} e_x \bar{e}_y} + \ket{a \bar{a} \bar{e}_x e_y} \right)$, respectively, where $a$, $e_x$, $e_y$ (spin-up) and $\bar{a}$, $\bar{e}_x$, $\bar{e}_y$ (spin-down) denote $a_1$ and $e$ orbitals. To obtain the $M_{S} = 0$ ground state, we used a closed-shell Hartree-Fock state $\ket{a \bar{a} e_x \bar{e}_x}$ as reference; the $M_{S} = 1$ ground state is itself an open-shell Hartree-Fock state, so we started with a higher energy reference state $\ket{a e_x \bar{e}_x e_y}$ in the $^3E$ manifold. We used unitary coupled-cluster single and double (UCCSD) ansatzs \cite{Peruzzo2014} to represent the trial wavefunctions. Fig. \ref{quantum}b and Fig. \ref{quantum}c show the estimated ground state energy as a function of the number of VQE iterations, where VQE calculations performed with the quantum simulator correctly converges to the ground state energy in both the $M_{S} = 1$ and $M_{S} = 0$ case. Despite the presence of noise, whose characterization and study will be critical to improve the use of quantum algorithms \cite{Kandala2019}, the results obtained with the quantum computer converge to the ground state energy within a 0.2 eV error. Calculations of excited states with quantum algorithms will be the focus of future works.

\section*{Discussion}

With the goal of providing a strategy to solve complex materials problems on NISQ computers, we proposed a first-principles quantum embedding theory where appropriate active regions of a material and their environment are described with different levels of accuracy, and the whole system is treated quantum mechanically. In particular, we used hybrid density functional theory for the environment, and we built a many-body Hamiltonian for the active space with effective electron-electron interactions that include dielectric screening and exchange-correlation effects from the environment. Our method overcomes the commonly used random phase approximation, which neglects exchange-correlation effects; importantly it is applicable to heterogeneous materials and scalable to large systems, thanks to the novel algorithms used here to compute response functions \cite{Ma2018,Nguyen2019}. We emphasize that the embedding theory presented here provides a flexible framework where multiple effects of the environment may be easily incorporated.  For instance, dynamical screening effects can be included by considering a frequency-dependent screened Coulomb interaction,  evaluated using the same procedure as the one outlined here for static screening; electron-phonon coupling effects can be incorporated by including phonon contributions in the screened Coulomb interactions. Furthermore, for systems where the electronic structure of the active region is expected to influence that of the host material, a self-consistent cycle in the calculation of the screened Coulomb interaction of the environment can be easily added to the approach.

We presented results for spin-defects in semiconductors obtained with both classical and quantum algorithms, and we showed excellent agreement between the two sets of techniques. Importantly, for selected cases we showed results obtained using a quantum simulator and a quantum computer, which agree within a relatively small error, in spite of the presence of noise in the quantum hardware. We predicted several excited states not observed before, in particular our results for the SiV in diamond and Cr in SiC represent the very first theoretical predictions of their singlet states, and provide important insights into their full optical cycle. We also demonstrated that a treatment of the dielectric screening beyond the random phase approximation leads to an accurate prediction of excitation energies.

The method proposed in our work enables calculations of realistic, heterogeneous materials using the resources of NISQ computers. We demonstrated quantum simulations of strongly-correlated electronic states in considerably larger systems (with hundreds of atoms) than previous studies combining quantum simulation and quantum embedding \cite{Rubin2016,Yamazaki2018,Bauer2016,Kreula2016,Rungger2019}. We have studied solids with defects, not just pristine materials, which are of great interest for quantum technologies. The strategy adopted here is general and may be applied to a variety of problems, including the simulation of active regions in molecules and materials for the understanding and discovery of catalysts and new drugs, and of aqueous solutions containing complex dissolved species. We finally note that our approach is not restricted to strongly-correlated active regions and will be useful also in the case of weakly correlated systems, where different regions of a material may be treated with varying levels of accuracy. Hence we expect the strategy presented here to be widely applicable to carry out quantum simulations of materials on near-term quantum computers.

\section*{Methods}

\noindent \textbf{Density Functional Theory}

\noindent All ground state DFT calculations are performed with the Quantum Espresso code \cite{Giannozzi2009} using the plane-wave pseudopotential formalism. Electron-ion interactions are modeled with norm-conserving pseudopotentials from the SG15 library \cite{Schlipf2015}. A kinetic energy cutoff of 50 Ry is used. All geometries are relaxed with spin-unrestricted DFT calculations using the Perdew–Burke-Ernzerhof (PBE) functional \cite{Perdew1996} until forces acting on atoms are smaller than 0.013 $\mathrm{eV} / \mathrm{\AA}$. NV and SiV in diamond are modeled with 216-atom supercells; Cr in 4H-SiC is modeled with a 128-atom supercell. The Brillouin zone is sampled with the $\Gamma$ point.

\hspace{1cm}

\noindent \textbf{Construction of effective Hamiltonians}

\noindent Construction of effective Hamiltonians is performed with the WEST code \cite{Govoni2015}, starting from wavefunctions of spin-restricted DFT calculations. For this step, we remark that the use of hybrid functional is important for an accurate mean-field description of defect levels, even though the geometry of defects are well represented at the PBE level. We used a dielectric dependent hybrid (DDH) functional \cite{Skone2014} which self-consistently determines the fraction of exact exchange based on the dielectric constant of the host material. The DDH functional was shown to yield accurate band gaps of diamond and silicon carbide \cite{Skone2014}, as well as optical properties of defects \cite{Seo2016,Seo2017,Pham2017,Smart2018,Gerosa2018}. After hybrid functional solutions of the Kohn-Sham equations are obtained, iterative diagonalizations of $\chi_0$ are performed, and density response functions and $f_{\mathrm{xc}}$ of the system are represented on a basis consisting of the first 512 eigenpotentials of $\chi_0$. Finite field calculations of $f_{\mathrm{xc}}$ are performed by coupling the WEST code with the Qbox \cite{Gygi2008} code. FCI calculations \cite{Knowles1984} on the effective Hamiltonian are carried out using the PySCF \cite{Sun2017} code.

\hspace{1cm}

\noindent \textbf{Quantum simulations}

\noindent In order to carry out quantum simulations, a minimum model of the NV center is constructed by applying the embedding theory with $a_1$ and $e$ orbitals beyond the RPA. 

In PEA simulations, the Jordan-Wigner transformation \cite{Jordan1928} is used to map the fermionic effective Hamiltonian to a qubit Hamiltonian, and Pauli operators with prefactors smaller than $10^{-6}$ a.u. are neglected to reduce the circuit depth, which results in less than $10^{-4}$ a.u. (0.003 eV) change in  eigenvalues. In order to achieve optimal precision, the Hamiltonian is scaled such that 0 and 2.5 eV are mapped to phases $\phi = 0$ and $\phi = 1$ of the ancilla qubits, respectively. We used the first-order Trotter formula to split time evolution operators into 4 time slices.

In VQE simulations, the parity transformation \cite{Bravyi2017} is adopted. For the simulation of the $M_{S} = 1$ state, the resulting qubit Hamiltonian acts on 4 qubits and there are 2 variational parameters in the UCCSD ansatz. For the simulation of the $M_{S} = 0$ state, we fixed the occupation of the $a$ orbital and the resulting qubit Hamiltonian acts on 2 qubits. We replicated the exponential excitation operator twice, with parameters in both replicas variationally optimized. Such a choice results in 6 variational parameters, providing a sufficient number of degrees of freedom for an accurate representation of the strongly-correlated $M_{S} = 0$ state. Parameters in the ansatz are optimized with the COBYLA algorithm \cite{Powell1994}.

Quantum simulations are performed with the QASM simulator and the IBM Q 5 Yorktown quantum computer using the IBM Qiskit package \cite{Qiskit}. Each quantum circuit is executed 8192 times to obtain statistically reliable sampling of the measurement results.

\pagebreak

\bibliography{ref}

\begin{thebibliography}{10}
\expandafter\ifx\csname url\endcsname\relax
  \def\url#1{\texttt{#1}}\fi
\expandafter\ifx\csname urlprefix\endcsname\relax\def\urlprefix{URL }\fi
\providecommand{\bibinfo}[2]{#2}
\providecommand{\eprint}[2][]{\url{#2}}

\bibitem{Cohen2008}
\bibinfo{author}{Cohen, A.~J.}, \bibinfo{author}{Mori-S{\'a}nchez, P.} \&
  \bibinfo{author}{Yang, W.}
\newblock \bibinfo{title}{Insights into current limitations of density
  functional theory}.
\newblock \emph{\bibinfo{journal}{Science}} \textbf{\bibinfo{volume}{321}},
  \bibinfo{pages}{792--794} (\bibinfo{year}{2008}).

\bibitem{Su2018}
\bibinfo{author}{Su, N.~Q.}, \bibinfo{author}{Li, C.} \& \bibinfo{author}{Yang,
  W.}
\newblock \bibinfo{title}{Describing strong correlation with fractional-spin
  correction in density functional theory}.
\newblock \emph{\bibinfo{journal}{Proc. Natl. Acad. Sci. U. S. A.}}
  \textbf{\bibinfo{volume}{115}}, \bibinfo{pages}{9678--9683}
  (\bibinfo{year}{2018}).

\bibitem{Georges1996}
\bibinfo{author}{Georges, A.}, \bibinfo{author}{Kotliar, G.},
  \bibinfo{author}{Krauth, W.} \& \bibinfo{author}{Rozenberg, M.~J.}
\newblock \bibinfo{title}{Dynamical mean-field theory of strongly correlated
  fermion systems and the limit of infinite dimensions}.
\newblock \emph{\bibinfo{journal}{Rev. Mod. Phys.}}
  \textbf{\bibinfo{volume}{68}}, \bibinfo{pages}{13--125}
  (\bibinfo{year}{1996}).

\bibitem{Kotliar2006}
\bibinfo{author}{Kotliar, G.} \emph{et~al.}
\newblock \bibinfo{title}{Electronic structure calculations with dynamical
  mean-field theory}.
\newblock \emph{\bibinfo{journal}{Rev. Mod. Phys.}}
  \textbf{\bibinfo{volume}{78}}, \bibinfo{pages}{865--951}
  (\bibinfo{year}{2006}).

\bibitem{Ceperley1986}
\bibinfo{author}{Ceperley, D.} \& \bibinfo{author}{Alder, B.}
\newblock \bibinfo{title}{Quantum monte carlo}.
\newblock \emph{\bibinfo{journal}{Science}} \textbf{\bibinfo{volume}{231}},
  \bibinfo{pages}{555--560} (\bibinfo{year}{1986}).

\bibitem{Wagner2016}
\bibinfo{author}{Wagner, L.~K.} \& \bibinfo{author}{Ceperley, D.~M.}
\newblock \bibinfo{title}{Discovering correlated fermions using quantum monte
  carlo}.
\newblock \emph{\bibinfo{journal}{Rep. Prog. Phys.}}
  \textbf{\bibinfo{volume}{79}}, \bibinfo{pages}{094501}
  (\bibinfo{year}{2016}).

\bibitem{Sun2017}
\bibinfo{author}{Sun, Q.} \emph{et~al.}
\newblock \bibinfo{title}{Py {SCF}: the python-based simulations of chemistry
  framework}.
\newblock \emph{\bibinfo{journal}{Wiley Interdiscip. Rev.: Comput. Mol. Sci.}}
  \textbf{\bibinfo{volume}{8}}, \bibinfo{pages}{e1340} (\bibinfo{year}{2017}).

\bibitem{Aspuru-Guzik2005}
\bibinfo{author}{Aspuru-Guzik, A.}
\newblock \bibinfo{title}{Simulated quantum computation of molecular energies}.
\newblock \emph{\bibinfo{journal}{Science}} \textbf{\bibinfo{volume}{309}},
  \bibinfo{pages}{1704--1707} (\bibinfo{year}{2005}).

\bibitem{Bravyi2017}
\bibinfo{author}{Bravyi, S.}, \bibinfo{author}{Gambetta, J.~M.},
  \bibinfo{author}{Mezzacapo, A.} \& \bibinfo{author}{Temme, K.}
\newblock \bibinfo{title}{Tapering off qubits to simulate fermionic
  hamiltonians}.
\newblock \emph{\bibinfo{journal}{arXiv preprint arXiv:1701.08213}}
  (\bibinfo{year}{2017}).

\bibitem{Babbush2018}
\bibinfo{author}{Babbush, R.} \emph{et~al.}
\newblock \bibinfo{title}{Low-depth quantum simulation of materials}.
\newblock \emph{\bibinfo{journal}{Phys. Rev. X}} \textbf{\bibinfo{volume}{8}},
  \bibinfo{pages}{011044} (\bibinfo{year}{2018}).

\bibitem{Kivlichan2018}
\bibinfo{author}{Kivlichan, I.~D.} \emph{et~al.}
\newblock \bibinfo{title}{Quantum simulation of electronic structure with
  linear depth and connectivity}.
\newblock \emph{\bibinfo{journal}{Phys. Rev. Lett.}}
  \textbf{\bibinfo{volume}{120}}, \bibinfo{pages}{110501}
  (\bibinfo{year}{2018}).

\bibitem{Motta2019}
\bibinfo{author}{Motta, M.} \emph{et~al.}
\newblock \bibinfo{title}{Determining eigenstates and thermal states on a
  quantum computer using quantum imaginary time evolution}.
\newblock \emph{\bibinfo{journal}{Nat. Phys.}} \bibinfo{pages}{1--6}
  (\bibinfo{year}{2019}).

\bibitem{Ollitrault2019}
\bibinfo{author}{Ollitrault, P.~J.} \emph{et~al.}
\newblock \bibinfo{title}{Quantum equation of motion for computing molecular
  excitation energies on a noisy quantum processor}.
\newblock \emph{\bibinfo{journal}{arXiv preprint arXiv:1910.12890}}
  (\bibinfo{year}{2019}).

\bibitem{Cao2019}
\bibinfo{author}{Cao, Y.} \emph{et~al.}
\newblock \bibinfo{title}{Quantum chemistry in the age of quantum computing}.
\newblock \emph{\bibinfo{journal}{Chem. Rev.}} \textbf{\bibinfo{volume}{119}},
  \bibinfo{pages}{10856--10915} (\bibinfo{year}{2019}).

\bibitem{Bauer2020}
\bibinfo{author}{Bauer, B.}, \bibinfo{author}{Bravyi, S.},
  \bibinfo{author}{Motta, M.} \& \bibinfo{author}{Chan, G. K.-L.}
\newblock \bibinfo{title}{Quantum algorithms for quantum chemistry and quantum
  materials science}.
\newblock \emph{\bibinfo{journal}{arXiv preprint arXiv:2001.03685}}
  (\bibinfo{year}{2020}).

\bibitem{Preskill2018}
\bibinfo{author}{Preskill, J.}
\newblock \bibinfo{title}{Quantum {C}omputing in the {NISQ} era and beyond}.
\newblock \emph{\bibinfo{journal}{{Quantum}}} \textbf{\bibinfo{volume}{2}},
  \bibinfo{pages}{79} (\bibinfo{year}{2018}).

\bibitem{Rubin2016}
\bibinfo{author}{Rubin, N.~C.}
\newblock \bibinfo{title}{A hybrid classical/quantum approach for large-scale
  studies of quantum systems with density matrix embedding theory}.
\newblock \emph{\bibinfo{journal}{arXiv preprint arXiv:1610.06910}}
  (\bibinfo{year}{2016}).

\bibitem{Yamazaki2018}
\bibinfo{author}{Yamazaki, T.}, \bibinfo{author}{Matsuura, S.},
  \bibinfo{author}{Narimani, A.}, \bibinfo{author}{Saidmuradov, A.} \&
  \bibinfo{author}{Zaribafiyan, A.}
\newblock \bibinfo{title}{Towards the practical application of near-term
  quantum computers in quantum chemistry simulations: A problem decomposition
  approach}.
\newblock \emph{\bibinfo{journal}{arXiv preprint arXiv:1806.01305}}
  (\bibinfo{year}{2018}).

\bibitem{Bauer2016}
\bibinfo{author}{Bauer, B.}, \bibinfo{author}{Wecker, D.},
  \bibinfo{author}{Millis, A.~J.}, \bibinfo{author}{Hastings, M.~B.} \&
  \bibinfo{author}{Troyer, M.}
\newblock \bibinfo{title}{Hybrid quantum-classical approach to correlated
  materials}.
\newblock \emph{\bibinfo{journal}{Phys. Rev. X}} \textbf{\bibinfo{volume}{6}},
  \bibinfo{pages}{031045} (\bibinfo{year}{2016}).

\bibitem{Kreula2016}
\bibinfo{author}{Kreula, J.~M.} \emph{et~al.}
\newblock \bibinfo{title}{Few-qubit quantum-classical simulation of strongly
  correlated lattice fermions}.
\newblock \emph{\bibinfo{journal}{EPJ Quantum Technol.}}
  \textbf{\bibinfo{volume}{3}}, \bibinfo{pages}{1--19} (\bibinfo{year}{2016}).

\bibitem{Rungger2019}
\bibinfo{author}{Rungger, I.} \emph{et~al.}
\newblock \bibinfo{title}{Dynamical mean field theory algorithm and experiment
  on quantum computers}.
\newblock \emph{\bibinfo{journal}{arXiv preprint arXiv:1910.04735}}
  (\bibinfo{year}{2019}).

\bibitem{Davies1976}
\bibinfo{author}{Davies, G.} \& \bibinfo{author}{Hamer, M.~F.}
\newblock \bibinfo{title}{Optical studies of the 1.945 {eV} vibronic band in
  diamond}.
\newblock \emph{\bibinfo{journal}{Proc. R. Soc. A}}
  \textbf{\bibinfo{volume}{348}}, \bibinfo{pages}{285--298}
  (\bibinfo{year}{1976}).

\bibitem{Rogers2008}
\bibinfo{author}{Rogers, L.~J.}, \bibinfo{author}{Armstrong, S.},
  \bibinfo{author}{Sellars, M.~J.} \& \bibinfo{author}{Manson, N.~B.}
\newblock \bibinfo{title}{Infrared emission of the {NV} centre in diamond:
  Zeeman and uniaxial stress studies}.
\newblock \emph{\bibinfo{journal}{New J. Phys.}} \textbf{\bibinfo{volume}{10}},
  \bibinfo{pages}{103024} (\bibinfo{year}{2008}).

\bibitem{Doherty2011}
\bibinfo{author}{Doherty, M.~W.}, \bibinfo{author}{Manson, N.~B.},
  \bibinfo{author}{Delaney, P.} \& \bibinfo{author}{Hollenberg, L. C.~L.}
\newblock \bibinfo{title}{The negatively charged nitrogen-vacancy centre in
  diamond: the electronic solution}.
\newblock \emph{\bibinfo{journal}{New J. Phys.}} \textbf{\bibinfo{volume}{13}},
  \bibinfo{pages}{025019} (\bibinfo{year}{2011}).

\bibitem{Maze2011}
\bibinfo{author}{Maze, J.~R.} \emph{et~al.}
\newblock \bibinfo{title}{Properties of nitrogen-vacancy centers in diamond:
  the group theoretic approach}.
\newblock \emph{\bibinfo{journal}{New J. Phys.}} \textbf{\bibinfo{volume}{13}},
  \bibinfo{pages}{025025} (\bibinfo{year}{2011}).

\bibitem{Choi2012}
\bibinfo{author}{Choi, S.}, \bibinfo{author}{Jain, M.} \&
  \bibinfo{author}{Louie, S.~G.}
\newblock \bibinfo{title}{Mechanism for optical initialization of spin in
  {NV}-center in diamond}.
\newblock \emph{\bibinfo{journal}{Phys. Rev. B}} \textbf{\bibinfo{volume}{86}},
  \bibinfo{pages}{041202} (\bibinfo{year}{2012}).

\bibitem{Doherty2013}
\bibinfo{author}{Doherty, M.~W.} \emph{et~al.}
\newblock \bibinfo{title}{The nitrogen-vacancy colour centre in diamond}.
\newblock \emph{\bibinfo{journal}{Phys. Rep.}} \textbf{\bibinfo{volume}{528}},
  \bibinfo{pages}{1--45} (\bibinfo{year}{2013}).

\bibitem{Goldman2015}
\bibinfo{author}{Goldman, M.~L.} \emph{et~al.}
\newblock \bibinfo{title}{State-selective intersystem crossing in
  nitrogen-vacancy centers}.
\newblock \emph{\bibinfo{journal}{Phys. Rev. B}} \textbf{\bibinfo{volume}{91}},
  \bibinfo{pages}{165201} (\bibinfo{year}{2015}).

\bibitem{Haenens2011}
\bibinfo{author}{D$^\prime$Haenens-Johansson, U. F.~S.} \emph{et~al.}
\newblock \bibinfo{title}{Optical properties of the neutral silicon
  split-vacancy center in diamond}.
\newblock \emph{\bibinfo{journal}{Phys. Rev. B}} \textbf{\bibinfo{volume}{84}},
  \bibinfo{pages}{245208} (\bibinfo{year}{2011}).

\bibitem{Gali2013}
\bibinfo{author}{Gali, A.} \& \bibinfo{author}{Maze, J.~R.}
\newblock \bibinfo{title}{Ab initio study of the split silicon-vacancy defect
  in diamond: Electronic structure and related properties}.
\newblock \emph{\bibinfo{journal}{Phys. Rev. B}} \textbf{\bibinfo{volume}{88}},
  \bibinfo{pages}{235205} (\bibinfo{year}{2013}).

\bibitem{Green2017}
\bibinfo{author}{Green, B.~L.} \emph{et~al.}
\newblock \bibinfo{title}{Neutral silicon-vacancy center in diamond: Spin
  polarization and lifetimes}.
\newblock \emph{\bibinfo{journal}{Phys. Rev. Lett.}}
  \textbf{\bibinfo{volume}{119}}, \bibinfo{pages}{096402}
  (\bibinfo{year}{2017}).

\bibitem{Rose2018}
\bibinfo{author}{Rose, B.~C.} \emph{et~al.}
\newblock \bibinfo{title}{Observation of an environmentally insensitive
  solid-state spin defect in diamond}.
\newblock \emph{\bibinfo{journal}{Science}} \textbf{\bibinfo{volume}{361}},
  \bibinfo{pages}{60--63} (\bibinfo{year}{2018}).

\bibitem{Green2019}
\bibinfo{author}{Green, B.~L.} \emph{et~al.}
\newblock \bibinfo{title}{Electronic structure of the neutral silicon-vacancy
  center in diamond}.
\newblock \emph{\bibinfo{journal}{Phys. Rev. B}} \textbf{\bibinfo{volume}{99}},
  \bibinfo{pages}{161112} (\bibinfo{year}{2019}).

\bibitem{Thiering2019}
\bibinfo{author}{Thiering, G.} \& \bibinfo{author}{Gali, A.}
\newblock \bibinfo{title}{The (eg $\otimes$ eu) $\otimes$ eg product
  jahn{\textendash}teller effect in the neutral group-{IV} vacancy quantum bits
  in diamond}.
\newblock \emph{\bibinfo{journal}{npj Comput. Mater.}}
  \textbf{\bibinfo{volume}{5}}, \bibinfo{pages}{18} (\bibinfo{year}{2019}).

\bibitem{Son1999}
\bibinfo{author}{Son, N.~T.} \emph{et~al.}
\newblock \bibinfo{title}{Photoluminescence and zeeman effect in chromium-doped
  4h and 6h {SiC}}.
\newblock \emph{\bibinfo{journal}{J. Appl. Phys.}}
  \textbf{\bibinfo{volume}{86}}, \bibinfo{pages}{4348--4353}
  (\bibinfo{year}{1999}).

\bibitem{Koehl2017}
\bibinfo{author}{Koehl, W.~F.} \emph{et~al.}
\newblock \bibinfo{title}{Resonant optical spectroscopy and coherent control of
  $\mathrm{C}{\mathrm{r}}^{4+}$ spin ensembles in sic and gan}.
\newblock \emph{\bibinfo{journal}{Phys. Rev. B}} \textbf{\bibinfo{volume}{95}},
  \bibinfo{pages}{035207} (\bibinfo{year}{2017}).

\bibitem{Diler2019}
\bibinfo{author}{Diler, B.} \emph{et~al.}
\newblock \bibinfo{title}{Coherent control and high-fidelity readout of
  chromium ions in commercial silicon carbide}.
\newblock \emph{\bibinfo{journal}{arXiv preprint arXiv:1909.08778}}
  (\bibinfo{year}{2019}).

\bibitem{Weber2010}
\bibinfo{author}{Weber, J.~R.} \emph{et~al.}
\newblock \bibinfo{title}{Quantum computing with defects}.
\newblock \emph{\bibinfo{journal}{Proc. Natl. Acad. Sci. U. S. A.}}
  \textbf{\bibinfo{volume}{107}}, \bibinfo{pages}{8513--8518}
  (\bibinfo{year}{2010}).

\bibitem{Seo2016}
\bibinfo{author}{Seo, H.}, \bibinfo{author}{Govoni, M.} \&
  \bibinfo{author}{Galli, G.}
\newblock \bibinfo{title}{Design of defect spins in piezoelectric aluminum
  nitride for solid-state hybrid quantum technologies}.
\newblock \emph{\bibinfo{journal}{Sci. Rep.}} \textbf{\bibinfo{volume}{6}},
  \bibinfo{pages}{20803} (\bibinfo{year}{2016}).

\bibitem{Seo2017}
\bibinfo{author}{Seo, H.}, \bibinfo{author}{Ma, H.}, \bibinfo{author}{Govoni,
  M.} \& \bibinfo{author}{Galli, G.}
\newblock \bibinfo{title}{Designing defect-based qubit candidates in wide-gap
  binary semiconductors for solid-state quantum technologies}.
\newblock \emph{\bibinfo{journal}{Phys. Rev. Mater.}}
  \textbf{\bibinfo{volume}{1}}, \bibinfo{pages}{075002} (\bibinfo{year}{2017}).

\bibitem{Ivady2018}
\bibinfo{author}{Iv{\'{a}}dy, V.}, \bibinfo{author}{Abrikosov, I.~A.} \&
  \bibinfo{author}{Gali, A.}
\newblock \bibinfo{title}{First principles calculation of spin-related
  quantities for point defect qubit research}.
\newblock \emph{\bibinfo{journal}{npj Comput. Mater.}}
  \textbf{\bibinfo{volume}{4}} (\bibinfo{year}{2018}).

\bibitem{Dreyer2018}
\bibinfo{author}{Dreyer, C.~E.}, \bibinfo{author}{Alkauskas, A.},
  \bibinfo{author}{Lyons, J.~L.}, \bibinfo{author}{Janotti, A.} \&
  \bibinfo{author}{Van~de Walle, C.~G.}
\newblock \bibinfo{title}{First-principles calculations of point defects for
  quantum technologies}.
\newblock \emph{\bibinfo{journal}{Annu. Rev. Mater. Res.}}
  \textbf{\bibinfo{volume}{48}}, \bibinfo{pages}{1--26} (\bibinfo{year}{2018}).
\newblock \eprint{https://doi.org/10.1146/annurev-matsci-070317-124453}.

\bibitem{Anderson2019}
\bibinfo{author}{Anderson, C.~P.} \emph{et~al.}
\newblock \bibinfo{title}{Electrical and optical control of single spins
  integrated in scalable semiconductor devices}.
\newblock \emph{\bibinfo{journal}{Science}} \textbf{\bibinfo{volume}{366}},
  \bibinfo{pages}{1225--1230} (\bibinfo{year}{2019}).
\newblock
  \eprint{https://science.sciencemag.org/content/366/6470/1225.full.pdf}.

\bibitem{Qiskit}
\bibinfo{author}{Abraham, H.} \emph{et~al.}
\newblock \bibinfo{title}{Qiskit: An open-source framework for quantum
  computing} (\bibinfo{year}{2019}).

\bibitem{Yorktown}
\bibinfo{note}{5-qubit backend: IBM Q team, ``IBM Q 5 Yorktown backend
  specification V1.1.0'', (2018). Retrieved from
  https://quantum-computing.ibm.com}.

\bibitem{Abrams1997}
\bibinfo{author}{Abrams, D.~S.} \& \bibinfo{author}{Lloyd, S.}
\newblock \bibinfo{title}{Simulation of many-body fermi systems on a universal
  quantum computer}.
\newblock \emph{\bibinfo{journal}{Phys. Rev. Lett.}}
  \textbf{\bibinfo{volume}{79}}, \bibinfo{pages}{2586} (\bibinfo{year}{1997}).

\bibitem{Peruzzo2014}
\bibinfo{author}{Peruzzo, A.} \emph{et~al.}
\newblock \bibinfo{title}{A variational eigenvalue solver on a photonic quantum
  processor}.
\newblock \emph{\bibinfo{journal}{Nat. Commun.}} \textbf{\bibinfo{volume}{5}},
  \bibinfo{pages}{4213} (\bibinfo{year}{2014}).

\bibitem{McClean2016}
\bibinfo{author}{McClean, J.~R.}, \bibinfo{author}{Romero, J.},
  \bibinfo{author}{Babbush, R.} \& \bibinfo{author}{Aspuru-Guzik, A.}
\newblock \bibinfo{title}{The theory of variational hybrid quantum-classical
  algorithms}.
\newblock \emph{\bibinfo{journal}{New J. Phys.}} \textbf{\bibinfo{volume}{18}},
  \bibinfo{pages}{023023} (\bibinfo{year}{2016}).

\bibitem{Kandala2017}
\bibinfo{author}{Kandala, A.} \emph{et~al.}
\newblock \bibinfo{title}{Hardware-efficient variational quantum eigensolver
  for small molecules and quantum magnets}.
\newblock \emph{\bibinfo{journal}{Nature}} \textbf{\bibinfo{volume}{549}},
  \bibinfo{pages}{242--246} (\bibinfo{year}{2017}).

\bibitem{Sun2016}
\bibinfo{author}{Sun, Q.} \& \bibinfo{author}{Chan, G. K.-L.}
\newblock \bibinfo{title}{Quantum embedding theories}.
\newblock \emph{\bibinfo{journal}{Acc. Chem. Res.}}
  \textbf{\bibinfo{volume}{49}}, \bibinfo{pages}{2705--2712}
  (\bibinfo{year}{2016}).

\bibitem{Huang2011}
\bibinfo{author}{Huang, C.}, \bibinfo{author}{Pavone, M.} \&
  \bibinfo{author}{Carter, E.~A.}
\newblock \bibinfo{title}{Quantum mechanical embedding theory based on a unique
  embedding potential}.
\newblock \emph{\bibinfo{journal}{J. Chem. Phys.}}
  \textbf{\bibinfo{volume}{134}}, \bibinfo{pages}{154110}
  (\bibinfo{year}{2011}).

\bibitem{Goodpaster2014}
\bibinfo{author}{Goodpaster, J.~D.}, \bibinfo{author}{Barnes, T.~A.},
  \bibinfo{author}{Manby, F.~R.} \& \bibinfo{author}{Miller, T.~F.}
\newblock \bibinfo{title}{Accurate and systematically improvable density
  functional theory embedding for correlated wavefunctions}.
\newblock \emph{\bibinfo{journal}{J. Chem. Phys.}}
  \textbf{\bibinfo{volume}{140}}, \bibinfo{pages}{18A507}
  (\bibinfo{year}{2014}).

\bibitem{Jacob2014}
\bibinfo{author}{Jacob, C.~R.} \& \bibinfo{author}{Neugebauer, J.}
\newblock \bibinfo{title}{Subsystem density-functional theory}.
\newblock \emph{\bibinfo{journal}{Wiley Interdiscip. Rev.: Comput. Mol. Sci.}}
  \textbf{\bibinfo{volume}{4}}, \bibinfo{pages}{325--362}
  (\bibinfo{year}{2014}).

\bibitem{Genova2014}
\bibinfo{author}{Genova, A.}, \bibinfo{author}{Ceresoli, D.} \&
  \bibinfo{author}{Pavanello, M.}
\newblock \bibinfo{title}{Periodic subsystem density-functional theory}.
\newblock \emph{\bibinfo{journal}{J. Chem. Phys.}}
  \textbf{\bibinfo{volume}{141}}, \bibinfo{pages}{174101}
  (\bibinfo{year}{2014}).

\bibitem{Wen2019}
\bibinfo{author}{Wen, X.}, \bibinfo{author}{Graham, D.~S.},
  \bibinfo{author}{Chulhai, D.~V.} \& \bibinfo{author}{Goodpaster, J.~D.}
\newblock \bibinfo{title}{Absolutely localized projection-based embedding for
  excited states}.
\newblock \emph{\bibinfo{journal}{arXiv preprint arXiv:1909.12423}}
  (\bibinfo{year}{2019}).

\bibitem{Knizia2012}
\bibinfo{author}{Knizia, G.} \& \bibinfo{author}{Chan, G. K.-L.}
\newblock \bibinfo{title}{Density matrix embedding: A simple alternative to
  dynamical mean-field theory}.
\newblock \emph{\bibinfo{journal}{Phys. Rev. Lett.}}
  \textbf{\bibinfo{volume}{109}}, \bibinfo{pages}{186404}
  (\bibinfo{year}{2012}).

\bibitem{Wouters2016}
\bibinfo{author}{Wouters, S.}, \bibinfo{author}{Jim{\'{e}}nez-Hoyos, C.~A.},
  \bibinfo{author}{Sun, Q.} \& \bibinfo{author}{Chan, G. K.-L.}
\newblock \bibinfo{title}{A practical guide to density matrix embedding theory
  in quantum chemistry}.
\newblock \emph{\bibinfo{journal}{J. Chem. Theory Comput.}}
  \textbf{\bibinfo{volume}{12}}, \bibinfo{pages}{2706--2719}
  (\bibinfo{year}{2016}).

\bibitem{Pham2019}
\bibinfo{author}{Pham, H.~Q.}, \bibinfo{author}{Hermes, M.~R.} \&
  \bibinfo{author}{Gagliardi, L.}
\newblock \bibinfo{title}{Periodic electronic structure calculations with
  density matrix embedding theory}.
\newblock \emph{\bibinfo{journal}{arXiv preprint arXiv:1909.08783}}
  (\bibinfo{year}{2019}).

\bibitem{Nguyen2016}
\bibinfo{author}{Lan, T.~N.}, \bibinfo{author}{Kananenka, A.~A.} \&
  \bibinfo{author}{Zgid, D.}
\newblock \bibinfo{title}{Rigorous ab initio quantum embedding for quantum
  chemistry using green's function theory: Screened interaction, nonlocal
  self-energy relaxation, orbital basis, and chemical accuracy}.
\newblock \emph{\bibinfo{journal}{J. Chem. Theory Comput.}}
  \textbf{\bibinfo{volume}{12}}, \bibinfo{pages}{4856--4870}
  (\bibinfo{year}{2016}).

\bibitem{Dvorak2019}
\bibinfo{author}{Dvorak, M.} \& \bibinfo{author}{Rinke, P.}
\newblock \bibinfo{title}{Dynamical configuration interaction: Quantum
  embedding that combines wave functions and green's functions}.
\newblock \emph{\bibinfo{journal}{Phys. Rev. B}} \textbf{\bibinfo{volume}{99}},
  \bibinfo{pages}{115134} (\bibinfo{year}{2019}).

\bibitem{Zhu2019}
\bibinfo{author}{Zhu, T.}, \bibinfo{author}{Cui, Z.-H.} \&
  \bibinfo{author}{Chan, G. K.-L.}
\newblock \bibinfo{title}{Efficient formulation of ab initio quantum embedding
  in periodic systems: Dynamical mean-field theory}.
\newblock \emph{\bibinfo{journal}{J. Chem. Theory Comput.}}
  (\bibinfo{year}{2019}).

\bibitem{Cui2019}
\bibinfo{author}{Cui, Z.-H.}, \bibinfo{author}{Zhu, T.} \&
  \bibinfo{author}{Chan, G. K.-L.}
\newblock \bibinfo{title}{Efficient implementation of ab initio quantum
  embedding in periodic systems: Density matrix embedding theory}.
\newblock \emph{\bibinfo{journal}{J. Chem. Theory Comput.}}
  (\bibinfo{year}{2019}).

\bibitem{Aryasetiawan2004}
\bibinfo{author}{Aryasetiawan, F.} \emph{et~al.}
\newblock \bibinfo{title}{Frequency-dependent local interactions and low-energy
  effective models from electronic structure calculations}.
\newblock \emph{\bibinfo{journal}{Phys. Rev. B}} \textbf{\bibinfo{volume}{70}},
  \bibinfo{pages}{195104} (\bibinfo{year}{2004}).

\bibitem{Miyake2009}
\bibinfo{author}{Miyake, T.}, \bibinfo{author}{Aryasetiawan, F.} \&
  \bibinfo{author}{Imada, M.}
\newblock \bibinfo{title}{Ab initio procedure for constructing effective models
  of correlated materials with entangled band structure}.
\newblock \emph{\bibinfo{journal}{Phys. Rev. B}} \textbf{\bibinfo{volume}{80}},
  \bibinfo{pages}{155134} (\bibinfo{year}{2009}).

\bibitem{Hirayama2017}
\bibinfo{author}{Hirayama, M.}, \bibinfo{author}{Miyake, T.},
  \bibinfo{author}{Imada, M.} \& \bibinfo{author}{Biermann, S.}
\newblock \bibinfo{title}{Low-energy effective hamiltonians for correlated
  electron systems beyond density functional theory}.
\newblock \emph{\bibinfo{journal}{Phys. Rev. B}} \textbf{\bibinfo{volume}{96}},
  \bibinfo{pages}{075102} (\bibinfo{year}{2017}).

\bibitem{Wilson2008}
\bibinfo{author}{Wilson, H.~F.}, \bibinfo{author}{Gygi, F.} \&
  \bibinfo{author}{Galli, G.}
\newblock \bibinfo{title}{Efficient iterative method for calculations of
  dielectric matrices}.
\newblock \emph{\bibinfo{journal}{Phys. Rev. B}} \textbf{\bibinfo{volume}{78}},
  \bibinfo{pages}{113303} (\bibinfo{year}{2008}).

\bibitem{Govoni2015}
\bibinfo{author}{Govoni, M.} \& \bibinfo{author}{Galli, G.}
\newblock \bibinfo{title}{Large scale {GW} calculations}.
\newblock \emph{\bibinfo{journal}{J. Chem. Theory Comput.}}
  \textbf{\bibinfo{volume}{11}}, \bibinfo{pages}{2680--2696}
  (\bibinfo{year}{2015}).

\bibitem{Ma2018}
\bibinfo{author}{Ma, H.}, \bibinfo{author}{Govoni, M.}, \bibinfo{author}{Gygi,
  F.} \& \bibinfo{author}{Galli, G.}
\newblock \bibinfo{title}{A finite-field approach {forGWCalculations} beyond
  the random phase approximation}.
\newblock \emph{\bibinfo{journal}{J. Chem. Theory Comput.}}
  \textbf{\bibinfo{volume}{15}}, \bibinfo{pages}{154--164}
  (\bibinfo{year}{2018}).

\bibitem{Nguyen2019}
\bibinfo{author}{Nguyen, N.~L.}, \bibinfo{author}{Ma, H.},
  \bibinfo{author}{Govoni, M.}, \bibinfo{author}{Gygi, F.} \&
  \bibinfo{author}{Galli, G.}
\newblock \bibinfo{title}{Finite-field approach to solving the bethe-salpeter
  equation}.
\newblock \emph{\bibinfo{journal}{Phys. Rev. Lett.}}
  \textbf{\bibinfo{volume}{122}}, \bibinfo{pages}{237402}
  (\bibinfo{year}{2019}).

\bibitem{Bockstedte2018}
\bibinfo{author}{Bockstedte, M.}, \bibinfo{author}{Sch\"utz, F.},
  \bibinfo{author}{Garratt, T.}, \bibinfo{author}{Iv{\'{a}}dy, V.} \&
  \bibinfo{author}{Gali, A.}
\newblock \bibinfo{title}{Ab initio description of highly correlated states in
  defects for realizing quantum bits}.
\newblock \emph{\bibinfo{journal}{npj Quantum Mater.}}
  \textbf{\bibinfo{volume}{3}}, \bibinfo{pages}{31} (\bibinfo{year}{2018}).

\bibitem{Skone2014}
\bibinfo{author}{Skone, J.~H.}, \bibinfo{author}{Govoni, M.} \&
  \bibinfo{author}{Galli, G.}
\newblock \bibinfo{title}{Self-consistent hybrid functional for condensed
  systems}.
\newblock \emph{\bibinfo{journal}{Phys. Rev. B}} \textbf{\bibinfo{volume}{89}},
  \bibinfo{pages}{195112} (\bibinfo{year}{2014}).

\bibitem{Heitz1995}
\bibinfo{author}{Heitz, R.} \emph{et~al.}
\newblock \bibinfo{title}{Identification of the 1.19-{eV} luminescence in
  hexagonal {GaN}}.
\newblock \emph{\bibinfo{journal}{Phys. Rev. B}} \textbf{\bibinfo{volume}{52}},
  \bibinfo{pages}{16508--16515} (\bibinfo{year}{1995}).

\bibitem{Anisimov1997}
\bibinfo{author}{Anisimov, V.~I.}, \bibinfo{author}{Aryasetiawan, F.} \&
  \bibinfo{author}{Lichtenstein, A.}
\newblock \bibinfo{title}{First-principles calculations of the electronic
  structure and spectra of strongly correlated systems: the lda+ u method}.
\newblock \emph{\bibinfo{journal}{J. Phys.: Condens. Matter}}
  \textbf{\bibinfo{volume}{9}}, \bibinfo{pages}{767} (\bibinfo{year}{1997}).

\bibitem{Kandala2019}
\bibinfo{author}{Kandala, A.} \emph{et~al.}
\newblock \bibinfo{title}{Error mitigation extends the computational reach of a
  noisy quantum processor}.
\newblock \emph{\bibinfo{journal}{Nature}} \textbf{\bibinfo{volume}{567}},
  \bibinfo{pages}{491} (\bibinfo{year}{2019}).

\bibitem{Giannozzi2009}
\bibinfo{author}{Giannozzi, P.} \emph{et~al.}
\newblock \bibinfo{title}{{QUANTUM} {ESPRESSO}: a modular and open-source
  software project for quantum simulations of materials}.
\newblock \emph{\bibinfo{journal}{J. Phys.: Condens. Matter}}
  \textbf{\bibinfo{volume}{21}}, \bibinfo{pages}{395502}
  (\bibinfo{year}{2009}).

\bibitem{Schlipf2015}
\bibinfo{author}{Schlipf, M.} \& \bibinfo{author}{Gygi, F.}
\newblock \bibinfo{title}{Optimization algorithm for the generation of {ONCV}
  pseudopotentials}.
\newblock \emph{\bibinfo{journal}{Comput. Phys. Commun.}}
  \textbf{\bibinfo{volume}{196}}, \bibinfo{pages}{36--44}
  (\bibinfo{year}{2015}).

\bibitem{Perdew1996}
\bibinfo{author}{Perdew, J.~P.}, \bibinfo{author}{Burke, K.} \&
  \bibinfo{author}{Ernzerhof, M.}
\newblock \bibinfo{title}{Generalized gradient approximation made simple}.
\newblock \emph{\bibinfo{journal}{Phys. Rev. Lett.}}
  \textbf{\bibinfo{volume}{77}}, \bibinfo{pages}{3865--3868}
  (\bibinfo{year}{1996}).

\bibitem{Pham2017}
\bibinfo{author}{Pham, T.~A.} \emph{et~al.}
\newblock \bibinfo{title}{Electronic structure of aqueous solutions: Bridging
  the gap between theory and experiments}.
\newblock \emph{\bibinfo{journal}{Science Advances}}
  \textbf{\bibinfo{volume}{3}} (\bibinfo{year}{2017}).

\bibitem{Smart2018}
\bibinfo{author}{Smart, T.~J.}, \bibinfo{author}{Wu, F.},
  \bibinfo{author}{Govoni, M.} \& \bibinfo{author}{Ping, Y.}
\newblock \bibinfo{title}{Fundamental principles for calculating charged defect
  ionization energies in ultrathin two-dimensional materials}.
\newblock \emph{\bibinfo{journal}{Phys. Rev. Materials}}
  \textbf{\bibinfo{volume}{2}}, \bibinfo{pages}{124002} (\bibinfo{year}{2018}).

\bibitem{Gerosa2018}
\bibinfo{author}{Gerosa, M.}, \bibinfo{author}{Gygi, F.},
  \bibinfo{author}{Govoni, M.} \& \bibinfo{author}{Galli, G.}
\newblock \bibinfo{title}{The role of defects and excess surface charges at
  finite temperature for optimizing oxide photoabsorbers}.
\newblock \emph{\bibinfo{journal}{Nature Materials}}
  \textbf{\bibinfo{volume}{17}}, \bibinfo{pages}{1122--1127}
  (\bibinfo{year}{2018}).

\bibitem{Gygi2008}
\bibinfo{author}{Gygi, F.}
\newblock \bibinfo{title}{Architecture of qbox: A scalable first-principles
  molecular dynamics code}.
\newblock \emph{\bibinfo{journal}{IBM J. Res. Dev.}}
  \textbf{\bibinfo{volume}{52}}, \bibinfo{pages}{137--144}
  (\bibinfo{year}{2008}).

\bibitem{Knowles1984}
\bibinfo{author}{Knowles, P.} \& \bibinfo{author}{Handy, N.}
\newblock \bibinfo{title}{A new determinant-based full configuration
  interaction method}.
\newblock \emph{\bibinfo{journal}{Chem. Phys. Lett.}}
  \textbf{\bibinfo{volume}{111}}, \bibinfo{pages}{315--321}
  (\bibinfo{year}{1984}).

\bibitem{Jordan1928}
\bibinfo{author}{Jordan, P.} \& \bibinfo{author}{Wigner, E.~P.}
\newblock \bibinfo{title}{About the pauli exclusion principle}.
\newblock \emph{\bibinfo{journal}{Z. Phys.}} \textbf{\bibinfo{volume}{47}},
  \bibinfo{pages}{631--651} (\bibinfo{year}{1928}).

\bibitem{Powell1994}
\bibinfo{author}{Powell, M.~J.}
\newblock \bibinfo{title}{A direct search optimization method that models the
  objective and constraint functions by linear interpolation}.
\newblock In \emph{\bibinfo{booktitle}{Advances in optimization and numerical
  analysis}}, \bibinfo{pages}{51--67} (\bibinfo{publisher}{Springer},
  \bibinfo{year}{1994}).

\end{thebibliography}
\bibliographystyle{naturemag}

\pagebreak

\section*{Acknowledgments}
We thank 
C. P. Anderson,
D. D. Awschalom,
T. C. Berkelbach,
B. Diler,
S. Dong,
D. Freedman,
L. Gagliardi,
F. Gygi,
F. J. Heremans,
L. Jiang,
A. M. Lewis,
A. Mezzacapo, 
P. J. Mintun,
H. Seo,
S. Sullivan,
S. J. Whiteley,
and G. Wolfowicz for fruitful discussions and comments on the manuscript.
We also thank the Qiskit Slack channel for generous help.
This work was supported by MICCoM, as part of the Computational Materials Sciences Program funded by the U.S. Department of Energy, Office of Science, Basic Energy Sciences, Materials Sciences and Engineering Division through Argonne National Laboratory, under contract number DE-AC02-06CH11357.
This research used resources of the National Energy Research Scientific Computing Center (NERSC), a DOE Office of Science User Facility supported by the Office of Science of the US Department of Energy under Contract No. DE-AC02-05CH11231, resources of the Argonne Leadership Computing Facility, which is a DOE Office of Science User Facility supported under Contract DE-AC02-06CH11357, and resources of the University of Chicago Research Computing Center.

\section*{Author contributions}
H.M., M.G., and G.G. designed the research. H.M. implemented the quantum embedding theory and performed simulations with classical and quantum algorithms, with supervision by M.G. and G.G. All authors wrote the manuscript.

\section*{Competing interests}
Authors declare no competing interests.

\pagebreak

\section*{Figures and tables}

\begin{table*}[ht]
\caption{Vertical excitation energies (eV) of the negatively charged nitrogen vacancy (NV) and neutral silicon vacancy (SiV) in diamond and Cr (4+) in 4H-SiC, obtained using the random phase approximation (RPA: third column) and including exchange-correlation interactions (beyond RPA: fourth column). Experimental measurements of zero-phonon-line (ZPL) energies are shown in brackets in the fifth column. Reference vertical excitation energies are computed from experimental ZPL when Stokes energies are available. }
\begin{center}
    \begin{tabular}{llccc}
    \hline
    System & Excitation & RPA & Beyond-RPA & Expt. \\
    \hline
    NV & ${}^3A_{2} \leftrightarrow {}^3E_{}$ &  1.921 &  2.001 &  2.180$^*$ (1.945$^*$) \\
    & ${}^3A_{2} \leftrightarrow {}^1A_{1}$ &  1.376 &  1.759 &                \\
    &  ${}^3A_{2} \leftrightarrow {}^1E_{}$ &  0.476 &  0.561 &                \\
    &  ${}^1E_{} \leftrightarrow {}^1A_{1}$ &  0.900 &  1.198 &        (1.190$^\dagger$) \\
    &  ${}^1A_{1} \leftrightarrow {}^3E_{}$ &  0.545 &  0.243 &  (0.344-0.430$^\ddagger$) \\
    \hline
    SiV &   ${}^3A_{2g} \leftrightarrow {}^3E_{u}$ &  1.590 &  1.594 &  1.568$^\mathsection$ (1.31$^{||}$) \\
    & ${}^3A_{2g} \leftrightarrow {}^3A_{1u}$ &  1.741 &  1.792 &               \\
    &  ${}^3A_{2g} \leftrightarrow {}^1E_{g}$ &  0.261 &  0.336 &               \\
    & ${}^3A_{2g} \leftrightarrow {}^1A_{1g}$ &  0.466 &  0.583 &               \\
    & ${}^3A_{2g} \leftrightarrow {}^1A_{1u}$ &  1.608 &  1.623 &               \\
    & ${}^3A_{2g} \leftrightarrow {}^1E_{u}$ &  2.056 &  2.171 &               \\
    & ${}^3A_{2g} \leftrightarrow {}^1A_{2u}$ &  2.365 &  2.515 &               \\
    &  ${}^3A_{2u} \leftrightarrow {}^3E_{u}$ &  0.003 &  0.011 &       (0.007$^{||}$) \\
    \hline
    Cr &   ${}^3A_{2} \leftrightarrow {}^3E_{}$ &  1.365 &  1.304 &          \\
   & ${}^3A_{2} \leftrightarrow {}^3A_{1}$ &  1.480 &  1.406 &          \\
   & ${}^3A_{2} \leftrightarrow {}^3E_{}'$ &  1.597 &  1.704 &          \\
   & ${}^3A_{2} \leftrightarrow {}^3A_{2}'$ &  1.635 &  1.755 &          \\
   & ${}^3A_{2} \leftrightarrow {}^1E_{}$ &  0.860 &  1.090 &  (1.190$^\mathparagraph$) \\
   & ${}^3A_{2} \leftrightarrow {}^1A_{1}$ &  1.560 &  1.937 &          \\
    \hline
    \end{tabular}
\label{excitation_energies}

$^*$Ref \cite{Davies1976}.
$^\dagger$Ref \cite{Rogers2008}.
$^\ddagger$Estimated by Ref \cite{Goldman2015} with a model for intersystem crossing.
$^\mathsection$Computed with Stokes energy from Ref \cite{Thiering2019}.
$^{||}$Ref \cite{Green2019}.
$^\mathparagraph$Ref \cite{Son1999}.
\end{center}
\end{table*}

\end{document}